\documentclass[conference,a4paper]{IEEEtran}

\usepackage{cite}
\usepackage{amsmath,amssymb,amsfonts}
\usepackage{algorithmic}
\usepackage{graphicx}
\usepackage{textcomp}
\usepackage{xcolor}
\def\BibTeX{{\rm B\kern-.05em{\sc i\kern-.025em b}\kern-.08em
    T\kern-.1667em\lower.7ex\hbox{E}\kern-.125emX}}

\let\savedCaption=\caption
\renewcommand*{\caption}[1]{%
	\vspace{-0.25cm}
	\savedCaption{#1}
	\vspace{-0.5cm}
}

\begin{document}

\title{Automated Mobile Video Objective Testing System}

\author{
	\IEEEauthorblockN{
		Eric Petajan\IEEEauthorrefmark{1},
		Jonathan Lynam\IEEEauthorrefmark{2}, 
		Morey Antebi\IEEEauthorrefmark{1},  
		Hessam Moeini\IEEEauthorrefmark{5},\\ 
		David Lindero\IEEEauthorrefmark{3}, 
		Lars Ernstrom\IEEEauthorrefmark{2}, 
		Gyanesh Patra\IEEEauthorrefmark{4}, 
		Szilveszter Nadas\IEEEauthorrefmark{2}
	}
	\IEEEauthorblockA{
		\IEEEauthorrefmark{1}AT\&T, New York, USA;
		\IEEEauthorrefmark{5}AT\&T, Plano, Texas, USA;
		\IEEEauthorrefmark{2}Ericsson Research, Santa Clara, California, USA;\\
		\IEEEauthorrefmark{3}Ericsson Research, Lulea, Sweden;
		\IEEEauthorrefmark{4}Ericsson Research, Kista, Sweden\\
		Email: ep619a@att.com, jonathan.lynam@ericsson.com
	}
	\\[-5.8ex]
}

\maketitle

\begin{figure}[!b]
\noindent\footnotesize\raggedright
\copyright~2025 IEEE. Personal use of this material is permitted. Permission from IEEE must
be obtained for all other uses, in any current or future media, including
reprinting/republishing this material for advertising or promotional purposes, creating new
collective works, for resale or redistribution to servers or lists, or reuse of any
copyrighted component of this work in other works.
\par\smallskip
Accepted author manuscript. Published in: \emph{2025 17th International Conference on Quality
of Multimedia Experience (QoMEX)}, Madrid, Spain, September 2025, pp.~1--4.
DOI: 10.1109/QoMEX65720.2025.11219938
\end{figure}

\begin{abstract}
Applying QoE analysis to optimize usage of cellular spectrum is of high interest to mobile network operators. A key challenge is to be able to perform QoE measurement across very different types of apps, from DASH VoD to interactive applications such as Video Conferencing and Cloud Gaming. This paper presents AMVOTS, a QoE measurement system developed by AT\&T, which is flexible enough to support a large range of application types and network conditions. We also discuss using AMVOTS as part of a closed loop to prototype QoE-aware radio resource allocation.
\end{abstract}

\section{Introduction}

User demand for throughput on cellular networks grows year after year \cite{ericsson2025mobiletraffic}. By 2025, it exceeds 200 Exabytes per month (23GB / month for each human on earth), and will approach 500 exabytes by 2030. The physical laws governing radio signal propagation place limits on the supply of bandwidth, in contrast to the ever-growing and largely unconstrained demand. As a result, maximizing the value delivered to end users over licensed spectrum has become a key priority. To achieve this, a user-centric approach is emerging, focused on quantifying and improving the end-user experience across applications. We conceptualize this new goal as ``Happiness per Hertz''. Your phone's OS update can trickle down in the background, your video chat with your spouse should take precedence. 
More subtly, in a video conference application, the audio is usually more important than video of the slide presentation. 
Only the Radio Base Station (RBS) has a close view of the cellular congestion, so in the end it must be actively involved to arbitrate between flows, deciding the best tradeoffs between them, ideally guided by application information and user preferences shared with the network. The video QoE research community and Video Quality Experts Group (VQEG) in particular have laid out substantial groundwork. We believe the VQEG 5G Key Performance Indicators (5G KPI) project \cite{vqeg5gkpi} will be an all-important bridge between network operators and video experts as their roles begin to overlap.

\section{AMVOTS}

\begin{figure}[t]
	\centering
	\includegraphics[width=0.48\textwidth]{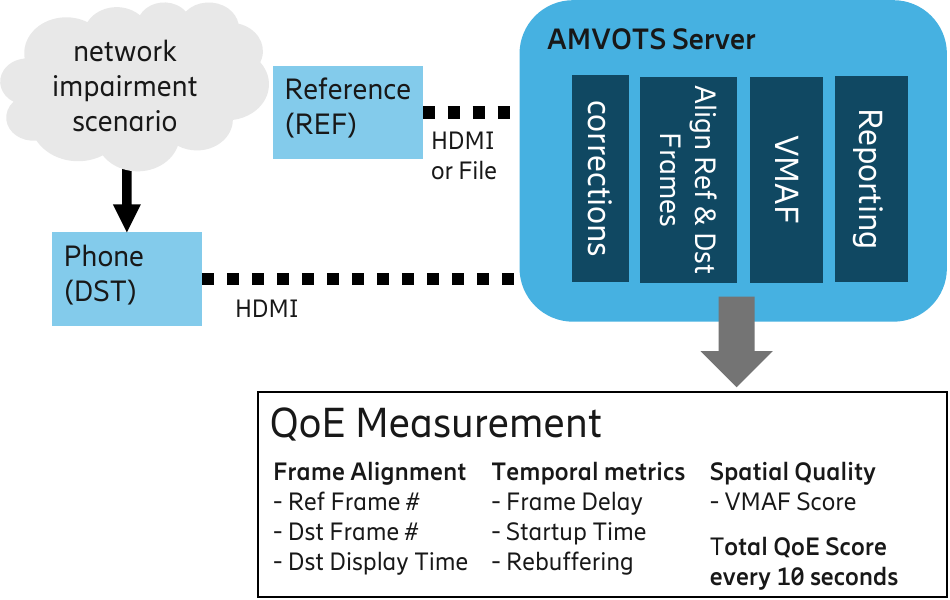}
	\caption{Architecture}
	\label{fig:arch}
\end{figure}

AMVOTS (Automated Mobile Video Objective Testing System) is a tool developed by AT\&T for performing lab-based video Quality of Experience (QoE) estimation. It fills a need to measure the video quality of closed-source video related applications, especially on mobile platforms, under various network conditions. This is of direct importance to mobile network operators (MNOs) in analyzing, predicting, and tuning spectrum usage. Utilizing VMAF \cite{vmaf1,vmaf2}, the system receives two video feeds, Reference and Distorted, matches up corresponding video frames between the video feeds and computes the VMAF score. Depending on the type of application being tested, the Reference and Distorted feeds can come from saved video clips or (more commonly) live HDMI capture, from computers or mobile phones (Fig. \ref{fig:arch}).

\begin{figure*}[t]
	\centering
	\includegraphics[width=0.75\textwidth]{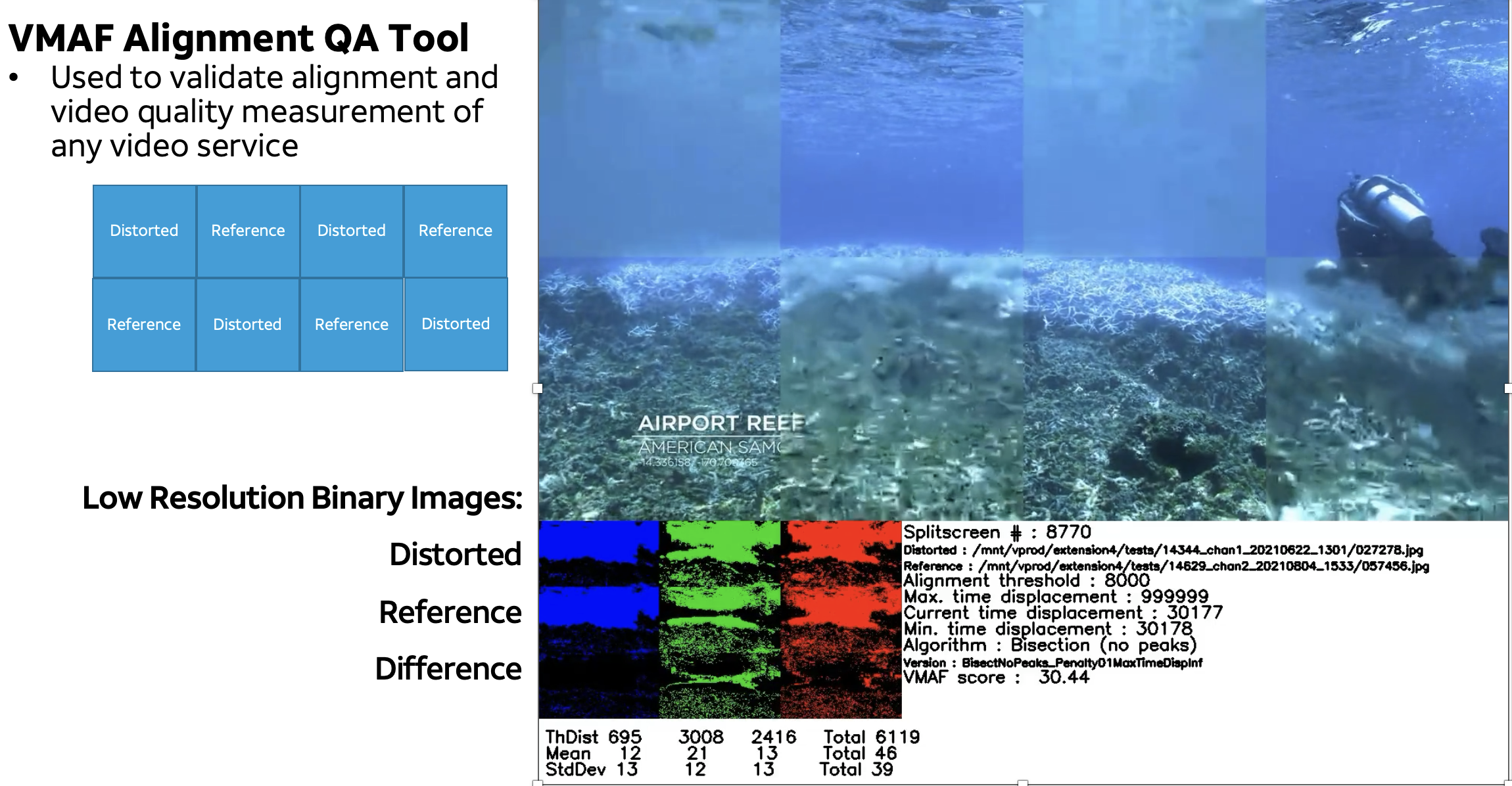}
	\caption{Realtime QoE Scoring}
	\label{fig:rt-scoring}
\end{figure*}

AMVOTS can be used to compute QoE scores for applications for which a Reference video feed (lossless or very high quality) can be obtained. Because it needs no integration beyond capturing video, it can perform QoE scoring even on closed-source applications. As a simple example, a low-latency remote desktop streaming application (such as Sunshine/Moonlight or Steam Link \cite{sun25,moon25,steaml25}) could provide a Reference feed from a PC (server), where the Distorted video feed is captured from a phone (client), where the video transport connecting them is subject to some network impairments. 

For testing Video on Demand (VoD) services, e.g. using DASH, the Reference feed could come from a pre-saved mezzanine video file (preferably), though if one is not available (e.g. a live news broadcast), the Reference could come from the top representation level offered by the VoD service. Ingesting saved files also allows for automation and repeatability of testing, for example, a video conference application can be tested repeatably by emulating a front-facing camera video input from a saved video file. That file would also be ingested as Reference, and Distorted feed would come from a UE participating in the video conference.

When both the Reference and Distorted feeds are live HDMI captures, it is possible to test interactive applications, including with live testing by humans. If Reference and Distorted feeds are from files, the result is similar to running native VMAF, but with the additional benefit that AMVOTS will automatically find the corresponding video frames between the files in a process known as {\it alignment}, even if the matching frames are at different time offsets or there are stalls or other timing differences. Also, various video corrections can be applied to normalize the video content between the feeds. The alignment algorithm and video corrections are discussed below.

AMVOTS processing server runs on a Dell R740 server with 28 cores, and an AJA Corvid 44 12G-SDI capture card. With that hardware it can capture a 1080p@60fps Reference/Distorted pair of video feeds, and run VMAF against the aligned frames. For more complex tests (e.g. multiple Distorted), it can capture up to 4 total feeds. It is possible to accept 4K inputs, but performance fluctuates and processing delays can occur. Since the system is primarily used for testing mobile phones, 1080p@60fps is usually sufficient, so less time has been spent optimizing the solution for higher resolutions.

\subsection{Normalizing and Correcting the Video Feeds}

In simple cases, an application plays video in full screen and its video feed can be ingested without changes. More often, though, video appears in sub-windows within a complex UI, e.g., in video conferencing, one participant might be shown prominently while others appear in smaller tiles. This variability requires cropping, scaling, rotation, or other normalization of the Reference and Distorted video feeds to produce accurate VMAF scores. The most common need is scaling. Another frequent issue arises when the Distorted side includes overlays, such as UI elements, logos, loading spinners, or pop-ups, that aren't present in the Reference. These must be masked before VMAF scoring by defining bounding boxes around affected regions. Simply filling these with black pixels would inflate VMAF scores, so instead AMVOTS copies image content from another user-specified bounding box, optionally applying a mirror transform. This yields more accurate results than black fill or ignoring the region. Finally, some apps, the OS or GPU, may apply image enhancements that alter brightness or color. To revert these enhancements, we added support for color correction using Hald \cite{jose2021haldclut}.

\begin{figure*}[t]
	\centering
	\includegraphics[width=0.85\textwidth]{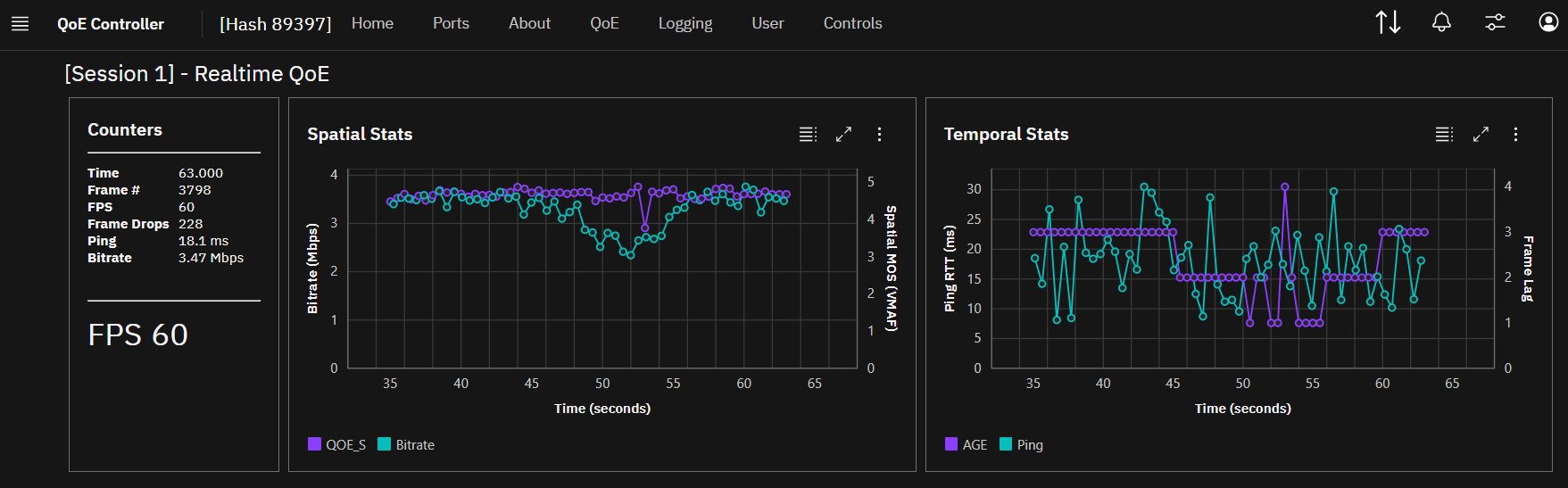}
	\caption{Realtime QoE dashboard}
	\label{fig:webui}
\end{figure*}

\subsection{Frame Alignment and Score Aggregation}

Before running VMAF, it is necessary to align the Reference and Distorted video frames, meaning that every Distorted frame is paired up with its corresponding Reference frame. This is non-trivial since the Distorted feed may involve frames that are delayed, missing (never shown), or duplicated. These problems can occur due to rebuffering (e.g. with DASH), network jitter, packet loss, or other causes. 

Since AMVOTS supports real-time QoE scoring, a high performance alignment algorithm is needed. The main part of the algorithm operates on a Reference, Distorted frame pair and computes a frame-difference (inverse of similarity) score. The score is computed by converting each frame into RGB24 color format, downscaled to 160x90 (for 16:9 content). Each of the R, G, B color channels are then separated into its own field and quantized to 1 bit, based on whether the pixel is above (1) or below (0) the mean intensity for that color field (in that frame). The field-level difference (from Reference and Distorted, for the same color) is the count of mismatching pixels. The per-frame difference is the sum of the field-level differences.

For efficiency, AMVOTS uses optimized CPU and GPU implementations of this algorithm, e.g. using XORs and POPCOUNT to sum the per color field differences. This frame-difference algorithm is run for each Distorted frame against a window of Reference frames whenever the two feeds appear to be out of sync. Some heuristics are applied to optimize for the normal case where the two feeds are in sync.

Figure \ref{fig:rt-scoring} shows the low resolution binary images per color channel for Distorted, Reference and their differences. Also shown in Figure \ref{fig:rt-scoring} is the alignment Quality Assurance (QA) view where alternating eighths of Distorted vs Reference easily show if the alignment isn't correct.

Prior to alignment, frame sequences in both the Reference and Distorted feeds are de-duplicated by computing a Mean Absolute Difference (MAD) between successive frames within a threshold. Also in this part of the system, source content freezes are detected (duplicate consecutive Reference frames), and disambiguated from stalls, which are duplicate Distorted frames paired to the same Reference frame. 

This algorithm is simple but very fast and effective in practice. The speed becomes important when searching across a large window of frames. It is accurate in aligning even in the presence of very low video quality; downscaling to 160x90 for the comparison reduces mismatches due to resolution differences, and matching color fields in a 1-bit quantization compensates for color space distortions.

Every 1 s, AMVOTS computes a Total QoE score data for the last 10 s. The frame VMAF scores are aggregated into a {\it video score} using harmonic mean. It also calculates an {\it interaction score} based on temporal factors, e.g., frame freezes and delays. We use a simplified aggregation method, until a standardized model by \cite{ itu_t_sg12} is available. The Total QoE for the 10 s interval is the minimum of these two scores. It is sent over the REST interface to the QoE controller (Fig. \ref{fig:webui}).

\section{QoE-In-The-Loop and QoE-Aware Resource Allocation}

AMVOTS  supports near-realtime QoE measurement and reporting. In this mode of operation, AMVOTS will send its frame alignment, timing, and VMAF score data over a REST interface to a server performing data collection. It is typically configured to report this data in small batches of 30 frames, the batching adds some latency to the reporting but helps improve execution overheads.

AMVOTS enables a form of testing we refer to as ``QoE-in-the-Loop'', where we can modify various aspects of a test scenario and analyze an application's dynamic behaviors and resulting impact on QoE. For example, in the Radio Access Network (RAN), analog signal attenuation equipment can be dynamically reconfigured to modify a phone's signal quality, or RBS traffic management parameters can be modified.

One major research direction in our work is QoE Aware Resource Allocation. This is where the RBS utilizes runtime QoE metrics to perform much more effective allocation of radio resources to each video flow to or from a UE. Our previous results \cite{nadas2024toqoe,nadas2025qoeaware} point to both QoE gains, and scalability gains. Often $\sim 3$ times more video flows can be supported at acceptable levels of Spatial QoE. Most of the gains are due to not wasting additional bandwidth on video feeds achieving almost perfect picture quality (e.g. VMAF 95+) but rather concentrating more bandwidth for flows that require more resources to achieve acceptable quality. 

To implement this technique in production would require QoE reporting either by applications or phone chipsets, which would require actions by multiple market players. By utilizing AMVOTS, we can prototype and experiment with such a system already. The QoE metrics are fed in real time to our ``QoE Controller'', a video-aware network function in our RAN testbed, which performs intelligent allocation of radio resources across the video flows. The QoE controller can operate either by modifying a flow’s traffic management parameters at the RBS, or for our own prototype video streaming applications, the QoE controller sends ``guidance'' messages, informing them of what bitrate they should set as their video encoding target in order to minimize queueing delays from traffic shaping at the RBS.

\section{Conclusion}

The creation and development of AMVOTS embodies AT\&T's commitment to delivering the best end-user experience to its cellular customers. The system is used to perform QoE scoring in lab settings, primarily on mobile phone applications tested on radio networks. Because it uses VMAF, a full-reference pixel based spatial QoE model, it allows testing of closed source applications requiring only access to HDMI video output from the phone. Video applications are the most bandwidth-intensive and among the most widely used services on cellular networks, making them a critical driver of both traffic volume and user experience. As mobile network operators begin to deploy products with increased awareness of video and QoE, we believe tools like AMVOTS will prove invaluable in helping vendors and operators make choices that bring the most value to end users.

\section*{Acknowledgment}

The authors gratefully acknowledge the extensive contributions of Michael Rose, who sadly passed away during this project.

\bibliographystyle{IEEEtran}
\bibliography{IEEEabrv,paper}

\begin{thebibliography}{10}
\providecommand{\url}[1]{#1}
\csname url@samestyle\endcsname
\providecommand{\newblock}{\relax}
\providecommand{\bibinfo}[2]{#2}
\providecommand{\BIBentrySTDinterwordspacing}{\spaceskip=0pt\relax}
\providecommand{\BIBentryALTinterwordstretchfactor}{4}
\providecommand{\BIBentryALTinterwordspacing}{\spaceskip=\fontdimen2\font plus
\BIBentryALTinterwordstretchfactor\fontdimen3\font minus
  \fontdimen4\font\relax}
\providecommand{\BIBforeignlanguage}[2]{{%
\expandafter\ifx\csname l@#1\endcsname\relax
\typeout{** WARNING: IEEEtran.bst: No hyphenation pattern has been}%
\typeout{** loaded for the language `#1'. Using the pattern for}%
\typeout{** the default language instead.}%
\else
\language=\csname l@#1\endcsname
\fi
#2}}
\providecommand{\BIBdecl}{\relax}
\BIBdecl

\bibitem{ericsson2025mobiletraffic}
\BIBentryALTinterwordspacing
(2025) {Mobile data traffic forecast – Ericsson Mobility Report}. Ericsson.
  Accessed: 2025-06-13. [Online]. Available:
  \url{https://www.ericsson.com/en/reports-and-papers/mobility-report/dataforecasts/mobile-traffic-forecast}
\BIBentrySTDinterwordspacing

\bibitem{vqeg5gkpi}
\BIBentryALTinterwordspacing
(2025) {5G Key Performance Indicators (5G KPI) - VQEG}. Video Quality Experts
  Group (VQEG). Accessed: 2025-06-13. [Online]. Available:
  \url{https://vqeg.org/projects/5gkpi/}
\BIBentrySTDinterwordspacing

\bibitem{vmaf1}
{Zhi Li, Christos Bampis, Julie Novak, Anne Aaron, Kyle Swanson, Anush Moorthy,
  and JD Cock}, ``Vmaf: The journey continues,''
  \url{https://netflixtechblog.com/vmaf-the-journey-continues-44b51ee9ed12},
  2018.

\bibitem{vmaf2}
{Zhi Li, Kyle Swanson, Christos Bampis, Lukáš Krasula and Anne Aaron},
  ``Toward a better quality metric for the video community, netflix technology
  blog,'' \url{https:
  //netflixtechblog.com/toward-a-better-quality-metric-for-the-video-community\
  -7ed94e752a30}, 2020.

\bibitem{sun25}
\BIBentryALTinterwordspacing
(2025) Sunshine. Accessed: 2025-06-11. [Online]. Available:
  \url{https://sunshine-stream.org}
\BIBentrySTDinterwordspacing

\bibitem{moon25}
\BIBentryALTinterwordspacing
(2025) Moonlight. Accessed: 2025-06-11. [Online]. Available:
  \url{https://moonlight-stream.org}
\BIBentrySTDinterwordspacing

\bibitem{steaml25}
\BIBentryALTinterwordspacing
(2025) Steam link. Accessed: 2025-06-11. [Online]. Available:
  \url{https://store.steampowered.com/steamlink}
\BIBentrySTDinterwordspacing

\bibitem{jose2021haldclut}
\BIBentryALTinterwordspacing
K.~M. Jose. (2021) Film simulations from scratch using python. Accessed:
  2025-06-13. [Online]. Available:
  \url{https://kevinmartinjose.com/2021/04/27/film-simulations-from-scratch-using-python/}
\BIBentrySTDinterwordspacing

\bibitem{itu_t_sg12}
``{ITU-T Study Group 12 (SG12): Performance, quality of service (QoS) and
  quality of experience (QoE)},''
  \url{https://www.itu.int/en/ITU-T/studygroups/2022-2024/12/Pages/default.aspx},
  {International Telecommunication Union - Telecommunication Standardization
  Sector (ITU-T)}, 2025, accessed: 2025-06-13.

\bibitem{nadas2024toqoe}
\BIBentryALTinterwordspacing
S.~N\'{a}das, L.~Ernstr\"{o}m, L.~Szil\'{a}gyi, G.~Patra, D.~Krylov, and
  J.~Lynam, ``{To QoE or not to QoE},'' in \emph{Proceedings of the 2024
  Applied Networking Research Workshop}, ser. ANRW '24.\hskip 1em plus 0.5em
  minus 0.4em\relax New York, NY, USA: Association for Computing Machinery,
  2024, p. 38–44. [Online]. Available:
  \url{https://doi.org/10.1145/3673422.3674892}
\BIBentrySTDinterwordspacing

\bibitem{nadas2025qoeaware}
\BIBentryALTinterwordspacing
S.~Nádas, L.~Ernström, D.~Lindero, and J.~Lynam, ``{On QoE-Aware Traffic
  Management for Real-time, Interactive Video with Time-variant Spatial
  Complexity},'' 2025. [Online]. Available:
  \url{https://arxiv.org/abs/2507.11798}
\BIBentrySTDinterwordspacing

\end{thebibliography}

\end{document}